\documentclass[10pt]{article}
\usepackage[latin1]{inputenc}
\usepackage{multicol}
\usepackage{amsmath,amssymb}
\usepackage[english]{babel}
\usepackage{layout}
\usepackage{enumerate}
\usepackage{hyperref}
\usepackage{color,graphicx,graphics}
\definecolor{green2}{rgb}{0,0.83,0}
\hypersetup{
            colorlinks={true},
            filecolor={0.19,0.27,0.69},
            urlcolor={blue},
            linkcolor={red},
            citecolor={green2},
            }
\newcommand\email[1]{{\tt\href{mailto:#1}{#1}}}
\numberwithin{equation}{section}
\begin{document}
\begin{center}\Large{\textsf{\textbf{Boundary Term in Metric $\boldsymbol{f(R)}$ Gravity: Field Equations in the Metric Formalism}}}\\
\end{center}
\begin{center}
Alejandro Guarnizo \footnote{\email{aguarnizot@unal.edu.co}}, Leonardo Casta\~neda \footnote{\email{lcastanedac@unal.edu.co}} \& Juan M. Tejeiro \footnote{\email{jmtejeiros@unal.edu.co}}
\end{center}
\begin{center}
\textit{Grupo de Gravitaci\'on y Cosmolog\'ia, Observatorio Astron\'omico Nacional,\\ Universidad Nacional de Colombia}\\
\textit{Bogot\'a-Colombia}
\end{center}
\date{}
\begin{abstract}
\noindent The main goal of this paper is to get in a straightforward form the field equations in metric $f(R)$ gravity, using elementary variational principles and adding a boundary term in the action, instead of the usual treatment in an equivalent scalar-tensor approach. We start with a brief review of the Einstein-Hilbert action, together with the Gibbons-York-Hawking boundary term, which is mentioned in some literature, but is generally missing. Next we present in detail the field equations in metric $f(R)$ gravity, including the discussion about boundaries, and we compare with the Gibbons-York-Hawking term in General Relativity. We notice that this boundary term is necessary in order to have a well defined extremal action principle under metric variation.
\end{abstract}
\section{Introduction}
General Relativity (GR) is the most widely accepted gravity theory proposed by Einstein in 1916, and it has been tested in several field strength regimes being on of the most successful and accurate theories in physics \cite{Will}. The field equations can be obtained using a variational principle, from the well know Einstein-Hilbert action \cite{Misner}-\cite{Weinberg}. The methodology leads to a boundary contribution which is usually dropped out \cite{Carroll},\cite{Hawking}, setting null fluxes through Gauss-Stokes theorem. It can be done by imposing that the variation of the metric and its first derivative vanishes in the boundary \cite{Wald}. These conditions can be relaxed whether a boundary term is introduced, called the Gibbons-York-Hawking boundary term \cite{Hawking1},\cite{Hawking2}. With this boundary term is necessary only to fix the variation of the metric in the boundary. There are some references \cite{Wald}-\cite{Padmanabhan} where this boundary term is shown explicitly. \\\\
However, GR is not the only relativistic theory of gravity. In the last decades several generalizations of Einstein field equations have been proposed \cite{Schmidt}-\cite{Querella}. Within these extended theories of gravity nowadays a subclass, known as $f(R)$ theories, are an alternative for classical problems, as the accelerated expansion of the universe, instead of Dark Energy and Quintessence models \cite{Nojiri}-\cite{Capozziello2}. $f(R)$ theories of gravity are basically extensions of the Einstein-Hilbert action with an arbitrary function of the Ricci scalar $R$ \cite{Faraoni}-\cite{Sotiriou}. The field equations were founded in \cite{Buchdahl}, and including boundary terms in fourth order gravity in \cite{Barth}-\cite{Odintsov2}. The Gibbons-York-Hawking like term in $f(R)$ gravity was explored in \cite{Madsen}, with an augmented variational principle in \cite{Fatibene},\cite{Francaviglia}, and using a scalar-tensor framework in \cite{Casadio}-\cite{Dyer}. Here we obtain the field equations from a metric $f(R)$ action with boundary terms, using only variational principles. We get a well constrained mathematical problem setting $\delta g_{\alpha\beta} = 0$ and $\delta R = 0$ in the boundary.
\section{General Relativity: The Einstein-Hilbert action with the Gibbons-York-Hawking boundary term}
We consider the space-time as a pair $(\mathcal{M},g)$ with $\mathcal{M}$ a four-dimensional manifold and $g$ a metric on $\mathcal{M}$. GR is based on the Einstein's Field equations (without cosmological constant and geometrical units $c=1$), which gives the form of the metric $g_{\alpha\beta}$ on the manifold $\mathcal{M}$:
\begin{equation}\label{campo1}
R_{\alpha\beta} - \frac{1}{2}R g_{\alpha\beta} = \kappa T_{\alpha\beta},
\end{equation}
where $R_{\alpha\beta}=R_{\alpha\eta\beta}^{\eta}$ is the Ricci tensor, $R=R^{\alpha\beta}R_{\alpha\beta}$ the Ricci scalar, and $T_{\alpha\beta}$ the stress-energy tensor, with $\kappa=8\pi G$, and sign convention $(-,+,+,+)$. The Riemann tensor is given by:
\begin{equation}
R_{\beta\gamma\delta}^{\alpha} = \partial_{\gamma}\Gamma_{\delta\beta}^{\alpha} - \partial_{\delta}\Gamma_{\gamma\beta}^{\alpha} + \Gamma_{\gamma\sigma}^{\alpha} \Gamma_{\delta\beta}^{\sigma} - \Gamma_{\sigma\delta}^{\alpha}\Gamma_{\gamma\beta}^{\sigma},
\end{equation}
in terms of the connections $\Gamma_{\beta\gamma}^{\alpha}$. The Einstein field equations can be recovered by using the variational principle $\delta S = 0$, with $S$ expressing the total action. In terms of Einstein-Hilbert action $S_{EH}$, Gibbons-York-Hawking boundary term $S_{GYH}$ and the action associated with all the matter fields $S_{M}$, the total action can be written by \cite{Poisson}:
\begin{equation}
S = \frac{1}{2\kappa}\bigl(S_{EH} + S_{GYH}\bigr) + S_{M},
\end{equation}
where
\begin{equation}\label{eh1}
S_{EH} = \int_{\mathcal{V}} d^4x\, \sqrt{-g}R,
\end{equation}
\begin{equation}
S_{GYH} = 2\oint_{\partial \mathcal{V}} d^3y \, \varepsilon\sqrt{|h|}K,
\end{equation}
here $\mathcal{V}$ is a hypervolume on $\mathcal{M}$, $\partial \mathcal{V}$ its boundary, $h$ the determinant of the induced metric, $K$ is the trace of the extrinsic curvature of the boundary $\partial \mathcal{V}$, and $\varepsilon$ is equal to $+1$ if $\partial \mathcal{V}$ is timelike and $-1$ if $\partial \mathcal{V}$ is spacelike (it is assumed that $\partial \mathcal{V}$ is nowhere null). Coordinates $x^{\alpha}$ are used for the finite region $\mathcal{V}$ and $y^{\alpha}$ for the boundary $\partial \mathcal{V}$. Now we will obtain the Einstein field equations varying the action with respect to $g^{\alpha\beta}$. We fixed the variation with the condition \cite{Wald},\cite{Poisson}
\begin{equation}\label{frontera}
\delta g_{\alpha\beta}\biggl|_{\partial \mathcal{V}} =0,
\end{equation}
i.e., the variation of the metric tensor vanishes in the boundary $\partial \mathcal{V}$. We use the results \cite{Poisson},\cite{Carroll}
\begin{equation}\label{varmet1}
\delta g_{\alpha\beta} = -g_{\alpha\mu}g_{\beta\nu}\delta g^{\mu\nu}, \qquad \delta g^{\alpha\beta} = -g^{\alpha\mu}g^{\beta\nu}\delta g_{\mu\nu},
\end{equation}
\begin{equation}
\delta \sqrt{-g} =  -\frac{1}{2}\sqrt{-g} g_{\alpha\beta}\delta g^{\alpha\beta},
\end{equation}
\begin{equation}
\delta R_{\beta\gamma\delta}^{\alpha} = \nabla_{\gamma}(\delta\Gamma_{\delta\beta}^{\alpha}) - \nabla_{\delta}(\delta\Gamma_{\gamma\beta}^{\alpha}),
\end{equation}
\begin{equation}\label{palatini}
\delta R_{\alpha\beta} = \nabla_{\gamma}(\delta\Gamma_{\beta\alpha}^{\gamma}) - \nabla_{\beta}(\delta\Gamma_{\gamma\alpha}^{\gamma}).
\end{equation}
We give a detailed review for the variation principles in GR following \cite{Wald},\cite{Poisson} and \cite{Carroll},. The variation of the Einstein-Hilbert term gives
\begin{equation}\label{varaccion1}
\delta S_{EH} = \int_{\mathcal{V}} d^4x \, \bigl(R\delta\sqrt{-g} + \sqrt{-g}\, \delta R\bigr).
\end{equation}
Now with $R = g^{\alpha\beta}R_{\alpha\beta}$, we have that the variation of the Ricci scalar is
\begin{equation}
\delta R = \delta g^{\alpha\beta}R_{\alpha\beta} + g^{\alpha\beta}\delta R_{\alpha\beta}.
\end{equation}
using the Palatini's identity (\ref{palatini}) we can write \cite{Carroll}:
\begin{align}
\delta R &= \delta g^{\alpha\beta}R_{\alpha\beta} + g^{\alpha\beta}\bigl(\nabla_{\gamma}(\delta\Gamma_{\beta\alpha}^{\gamma}) - \nabla_{\beta}(\delta\Gamma_{\alpha\gamma}^{\gamma})\bigr),\nonumber \\
&= \delta g^{\alpha\beta}R_{\alpha\beta} + \nabla_{\sigma
}\bigl(g^{\alpha\beta}(\delta\Gamma_{\beta\alpha}^{\sigma}) - g^{\alpha\sigma}
(\delta\Gamma_{\alpha\gamma}^{\gamma})\bigr),
\end{align}
where we have used the metric compatibility $\nabla_{\gamma}g_{\alpha\beta}\equiv 0$ and relabeled some dummy indices. Inserting this results for the variations in expression (\ref{varaccion1}) we have:
\begin{align}
\delta S_{EH} &= \int_{\mathcal{V}} d^4x \, \bigl(R\delta\sqrt{-g} + \sqrt{-g}\, \delta R\bigr),\nonumber \\
&= \int_{\mathcal{V}} d^4x \, \biggl(-\frac{1}{2}Rg_{\alpha\beta}\sqrt{-g}\, \delta g^{\alpha\beta} + R_{\alpha\beta}\sqrt{-g}\delta g^{\alpha\beta} + \sqrt{-g}\nabla_{\sigma}\bigl(g^{\alpha\beta}(\delta\Gamma_{\beta\alpha}^{\sigma}) - g^{\alpha\sigma}
(\delta\Gamma_{\alpha\gamma}^{\gamma})\bigr)\biggr),\nonumber \\
&= \int_{\mathcal{V}}  d^4x \, \sqrt{-g} \biggl(R_{\alpha\beta}-\frac{1}{2}Rg_{\alpha\beta}\biggr)\delta g^{\alpha\beta} + \int_{\mathcal{V}}  d^4x\sqrt{-g} \nabla_{\sigma}\bigl(g^{\alpha\beta}(\delta\Gamma_{\beta\alpha}^{\sigma}) - g^{\alpha\sigma}
(\delta\Gamma_{\alpha\gamma}^{\gamma})\bigr).
\end{align}
Denoting the divergence term with $\delta S_{B}$,
\begin{equation}
\delta S_B = \int_{\mathcal{V}} d^4x \, \sqrt{-g}\,  \nabla_{\sigma}\bigl(g^{\alpha\beta}(\delta\Gamma_{\beta\alpha}^{\sigma}) - g^{\alpha\sigma}
(\delta\Gamma_{\alpha\gamma}^{\gamma})\bigr),
\end{equation}
we define
\begin{equation}\label{V}
V^{\sigma} = g^{\alpha\beta}(\delta\Gamma_{\beta\alpha}^{\sigma}) - g^{\alpha\sigma}
(\delta\Gamma_{\alpha\gamma}^{\gamma}),
\end{equation}
then the boundary term can be written as
\begin{equation}\label{boundary}
\delta S_B = \int_{\mathcal{V}} d^4x \, \sqrt{-g}\,  \nabla_{\sigma}V^{\sigma}.
\end{equation}
Using Gauss-Stokes theorem \cite{Poisson},\cite{Carroll}:
\begin{equation}\label{gauss}
\int_{\mathcal{V}} d^{n}x\, \sqrt{|g|}\nabla_{\mu}A^{\mu} = \oint_{\partial \mathcal{V}}d^{n-1}y\, \varepsilon\sqrt{|h|}n_{\mu}A^{\mu},
\end{equation}
where $n_{\mu}$ is the unit normal to $\partial \mathcal{V}$. Using this we can write (\ref{boundary}) in the following boundary term
\begin{equation}
\delta S_B = \oint_{\partial \mathcal{V}} d^{3}y\, \varepsilon \sqrt{|h|}n_{\sigma}V^{\sigma},
\end{equation}
with $V^{\sigma}$ given in (\ref{V}). The variation $\delta \Gamma_{\beta\alpha}^{\sigma}$ is obtained by using  that $\Gamma_{\beta\alpha}^{\sigma}$ is the Christoffel symbol $\bigl\{_{\beta\alpha}^{\sigma}\bigr\}$:
\begin{equation}\label{simbolo}
\Gamma_{\beta\gamma}^{\alpha} \equiv \Bigl\{_{\beta\gamma}^{\alpha}\Bigr\} = \frac{1}{2}g^{\alpha\sigma}\bigl[\partial_{\beta}g_{\sigma\gamma} + \partial_{\gamma}g_{\sigma\beta} - \partial_{\sigma}g_{\beta\gamma}\bigr],
\end{equation}
getting
\begin{align}\label{varsim1}
\delta \Gamma_{\beta\alpha}^{\sigma} &= \delta \biggl(\frac{1}{2}g^{\sigma\gamma}\bigl[\partial_{\beta}g_{\gamma\alpha} + \partial_{\alpha}g_{\gamma\beta} - \partial_{\gamma}g_{\beta\alpha}\bigr]\biggr),\nonumber \\
&= \frac{1}{2}\delta g^{\sigma\gamma}\bigl[\partial_{\beta}g_{\gamma\alpha} + \partial_{\alpha}g_{\gamma\beta} - \partial_{\gamma}g_{\beta\alpha}\bigr] + \frac{1}{2}g^{\sigma\gamma}\bigl[\partial_{\beta}(\delta g_{\gamma\alpha}) + \partial_{\alpha}(\delta g_{\gamma\beta}) -  \partial_{\gamma}(\delta g_{\beta\alpha})\bigr].
\end{align}
From the boundary conditions $\delta g_{\alpha\beta} = \delta g^{\alpha\beta}=0$ the variation (\ref{varsim1}) gives:
\begin{equation}
\delta \Gamma_{\beta\alpha}^{\sigma}\Bigl|_{\partial \mathcal{V}} = \frac{1}{2}g^{\sigma\gamma}\bigl[\partial_{\beta}(\delta g_{\gamma\alpha}) + \partial_{\alpha}(\delta g_{\gamma\beta}) -  \partial_{\gamma}(\delta g_{\beta\alpha})\bigr],
\end{equation}
and
\begin{equation}
V^{\mu}\Bigl|_{\partial \mathcal{V}} = g^{\alpha\beta}\biggl[\frac{1}{2}g^{\mu\gamma}\bigl[\partial_{\beta}(\delta g_{\gamma\alpha}) + \partial_{\alpha}(\delta g_{\gamma\beta}) -  \partial_{\gamma}(\delta g_{\beta\alpha})\bigr]\biggr] - g^{\alpha\mu}\biggl[\frac{1}{2}g^{\nu\gamma}\partial_{\alpha}(\delta g_{\nu\gamma}) \biggr],
\end{equation}
we can write
\begin{align}
V_{\sigma}\Bigl|_{\partial \mathcal{V}} = g_{\sigma\mu}V^{\mu}\Bigl|_{\partial \mathcal{V}} &= g_{\sigma\mu}g^{\alpha\beta}\biggl[\frac{1}{2}g^{\mu\gamma}\bigl[\partial_{\beta}(\delta g_{\gamma\alpha}) + \partial_{\alpha}(\delta g_{\gamma\beta}) -  \partial_{\gamma}(\delta g_{\beta\alpha})\bigr]\biggr] - g_{\sigma\mu}g^{\alpha\mu}\biggl[\frac{1}{2}g^{\nu\gamma}\partial_{\alpha}(\delta g_{\nu\gamma}) \biggr], \nonumber \\
&= \frac{1}{2}\delta_{\sigma}^{\gamma}g^{\alpha\beta}\bigl[\partial_{\beta}(\delta g_{\gamma\alpha}) + \partial_{\alpha}(\delta g_{\gamma\beta}) -  \partial_{\gamma}(\delta g_{\beta\alpha})\bigr] - \frac{1}{2}\delta_{\sigma}^{\alpha}g^{\nu\gamma}\bigl[\partial_{\alpha}(\delta g_{\nu\gamma}) \bigr],\nonumber \\
&= g^{\alpha\beta}\bigl[\partial_{\beta}(\delta g_{\sigma\alpha}) -  \partial_{\sigma}(\delta g_{\beta\alpha})\bigr].
\end{align}
We now evaluate the term $n^{\sigma}V_{\sigma}\bigl|_{\partial \mathcal{V}}$ by using for this that
\begin{equation}
g^{\alpha\beta} = h^{\alpha\beta} + \varepsilon n^{\alpha}n^{\beta},
\end{equation}
then
\begin{align}
n^{\sigma}V_{\sigma}\Bigl|_{\partial \mathcal{V}} &= n^{\sigma}(h^{\alpha\beta}+\varepsilon n^{\alpha}n^{\beta})[\partial_{\beta}(\delta g_{\sigma\alpha}) -  \partial_{\sigma}(\delta g_{\beta\alpha})], \nonumber \\
&= n^{\sigma}h^{\alpha\beta}[\partial_{\beta}(\delta g_{\sigma\alpha}) -  \partial_{\sigma}(\delta g_{\beta\alpha})],
\end{align}
where we use the antisymmetric part of $\varepsilon n^{\alpha}n^{\beta}$ with $\varepsilon = n^{\mu}n_{\mu}=\pm 1$. To the fact $\delta g_{\alpha\beta}=0$ in the boundary  we have $h^{\alpha\beta}\partial_{\beta}(\delta g_{\sigma\alpha})=0$ \cite{Poisson}. Finally we get
\begin{equation}
n^{\sigma}V_{\sigma}\Bigl|_{\partial \mathcal{V}} = -n^{\sigma}h^{\alpha\beta}\partial_{\sigma}(\delta g_{\beta\alpha}).
\end{equation}
Thus the variation of the Einstein-Hilbert term is:
\begin{equation}
\delta S_{EH} = \int_{\mathcal{V}}  d^4x \, \sqrt{-g} \biggl(R_{\alpha\beta}-\frac{1}{2}Rg_{\alpha\beta}\biggr)\delta g^{\alpha\beta}-\oint_{\partial \mathcal{V}} d^{3}y\, \varepsilon \sqrt{|h|} h^{\alpha\beta}\partial_{\sigma}(\delta g_{\beta\alpha})n^{\sigma}.
\end{equation}
Now we consider the variation of the Gibbons-York-Hawking boundary term:\\
\begin{equation}
\delta S_{GYH} = 2\oint_{\partial \mathcal{V}} d^3y\, \varepsilon\sqrt{|h|}\delta K.
\end{equation}
Using the definition of the  trace of extrinsic curvature \cite{Poisson}:
\begin{align}
K &= \nabla_{\alpha}n^{\alpha}, \nonumber \\
&= g^{\alpha\beta}\nabla_{\beta}n_{\alpha}, \nonumber \\
&= (h^{\alpha\beta}+\varepsilon n^{\alpha}n^{\beta})\nabla_{\beta}n_{\alpha}, \nonumber \\
&= h^{\alpha\beta}\nabla_{\beta}n_{\alpha}, \nonumber \\
&= h^{\alpha\beta}(\partial_{\beta}n_{\alpha}-\Gamma_{\beta\alpha}^{\gamma}n_{\gamma}),
\end{align}
the variation is
\begin{align}\label{deltaK}
\delta K &= -h^{\alpha\beta}\delta\Gamma_{\beta\alpha}^{\gamma}n_{\gamma}, \nonumber \\
&= -\frac{1}{2}h^{\alpha\beta}g^{\sigma\gamma}\bigl[\partial_{\beta}(\delta g_{\sigma\alpha}) + \partial_{\alpha}(\delta g_{\sigma\beta}) -  \partial_{\sigma}(\delta g_{\beta\alpha})\bigr]n_{\gamma}, \nonumber \\
&= -\frac{1}{2}h^{\alpha\beta}\bigl[\partial_{\beta}(\delta g_{\sigma\alpha}) + \partial_{\alpha}(\delta g_{\sigma\beta}) -  \partial_{\sigma}(\delta g_{\beta\alpha})\bigr]n^{\sigma}, \nonumber \\
&= \frac{1}{2}h^{\alpha\beta}\partial_{\sigma}(\delta g_{\beta\alpha})n^{\sigma}.
\end{align}
This comes from the variation $\delta\Gamma_{\beta\alpha}^{\gamma}$ evaluated in the boundary, and the fact that $h^{\alpha\beta}\partial_{\beta}(\delta g_{\sigma\alpha})=0$, $h^{\alpha\beta}\partial_{\alpha}(\delta g_{\sigma\beta})=0$. Then we have for the variation of the Gibbons-York-Hawking boundary term:
\begin{equation}
\delta S_{GYH} = \oint_{\partial \mathcal{V}} d^3y\, \varepsilon\sqrt{|h|}h^{\alpha\beta}\partial_{\sigma}(\delta g_{\beta\alpha})n^{\sigma}.
\end{equation}
We see that this term exactly cancel the boundary contribution of the Einstein-Hilbert term. Now, if we have a matter action defined by:
\begin{equation}\label{matteraction}
S_M = \int_{\mathcal{V}} d^4x\, \sqrt{-g} \mathcal{L}_M[g_{\alpha\beta},\psi],
\end{equation}
where $\psi$ denotes the matter fields. The variation of this action takes the form:
\begin{align}
\delta S_M &= \int_{\mathcal{V}} d^4x\, \delta(\sqrt{-g} \mathcal{L}_M),\nonumber \\
&= \int_{\mathcal{V}} d^4x\, \biggl(\frac{\partial \mathcal{L}_M}{\partial g^{\alpha\beta}}\delta g^{\alpha\beta}\sqrt{-g} + \mathcal{L}_M\delta\sqrt{-g}\biggr),\nonumber \\
&= \int_{\mathcal{V}} d^4x\,\sqrt{-g} \biggl(\frac{\partial \mathcal{L}_M}{\partial g^{\alpha\beta}} -\frac{1}{2} \mathcal{L}_Mg_{\alpha\beta}\biggr)\delta g^{\alpha\beta},
\end{align}
as usual, defining the stress-energy tensor by:
\begin{equation}\label{tensem}
T_{\alpha\beta} \equiv -2\frac{\partial \mathcal{L}_M}{\partial g^{\alpha\beta}} + \mathcal{L}_Mg_{\alpha\beta} = -\frac{2}{\sqrt{-g}}\frac{\delta S_M}{\delta g^{\alpha\beta}},
\end{equation}
then:
\begin{equation}\label{variacionener}
\delta S_M = -\frac{1}{2}\int_{\mathcal{V}} d^4x\,\sqrt{-g} T_{\alpha\beta}\delta g^{\alpha\beta},
\end{equation}
imposing the total variations remains invariant with respect to $\delta g^{\alpha\beta}$. Finally the equations are writing as:
\begin{equation}\label{campo1}
\frac{1}{\sqrt{-g}}\frac{\delta S}{\delta g^{\alpha\beta}} = 0,  \Longrightarrow R_{\alpha\beta} - \frac{1}{2}R g_{\alpha\beta} = \kappa T_{\alpha\beta} ,
\end{equation}
which corresponds to Einstein field equations in geometric units $c=1$.
\section{Field equations in $f(R)$ gravity}
As we mentioned above the modified theories of gravity have been studied in order to explain among the accelerated expansion of the universe. One of these theories is the modified $f(R)$ gravity which consists in add additional higher order terms of the Ricci scalar in the Einstein-Hilbert action \cite{Nojiri},\cite{Sami},\cite{Sotiriou}.  There are three versions of $f(R)$ gravity: Metric formalism, Palatini formalism and metric-affine formalism \cite{Sotiriou}. Here we focus only in the metric formalism; for a detailed deduction of field equations in  the Palatini and the metric-affine formalism see \cite{Sotiriou3},\cite{Sotiriou4}.  Again, we consider the space-time as a pair $(\mathcal{M},g)$ with $\mathcal{M}$ a four-dimensional manifold and $g_{\alpha\beta}$ a metric on $\mathcal{M}$. Now the lagrangian is an arbitrary function of the Ricci scalar $\mathcal{L}[g_{\alpha\beta}] = f(R)$, the relation of the Ricci scalar and the metric tensor is given assuming a Levi-Civita connection of the manifold. i.e. a Christoffel symbol. This lagrangian was presented in \cite{Fatibene} using augmented variational principles. The general action can be written as \cite{Dyer}:
\begin{equation}
S_{mod} = \frac{1}{2\kappa}\bigl(S_{met} + S'_{GYH}\bigr) + S_{M},
\end{equation}
with the bulk term
\begin{equation} \label{accion1}
S_{met} = \int_{\mathcal{V}} d^4x\, \sqrt{-g}f(R),
\end{equation}\\
and the Gibbons-York-Hawking like boundary term \cite{Madsen},\cite{Dyer}
\begin{equation}
S'_{GYH} = 2\oint_{\partial \mathcal{V}} d^3y\, \varepsilon\sqrt{|h|} f'(R)K,
\end{equation}
with $f'(R) = df(R)/dR$. Again, $S_M$ represents the action associated with all the matter fields (\ref{matteraction}). We fixed the variation to the condition
\begin{equation}\label{frontera2}
\delta g_{\alpha\beta}\biggl|_{\partial \mathcal{V}} =0.
\end{equation}
First, the variation of the bulk term is:
\begin{equation}\label{varaccion2}
\delta S_{met} = \int_{\mathcal{V}} d^4x \, \bigl(f(R)\delta\sqrt{-g} + \sqrt{-g}\, \delta f(R)\bigr),
\end{equation}
and the functional derivative of the $f(R)$ term can be written as
\begin{equation}
\delta f(R) = f'(R) \delta R.
\end{equation}
Using the expression for the variation of the Ricci scalar:
\begin{equation}
\delta R = \delta g^{\alpha\beta}R_{\alpha\beta} + \nabla_{\sigma}\bigl(g^{\alpha\beta}(\delta\Gamma_{\beta\alpha}^{\sigma}) - g^{\alpha\sigma}
(\delta\Gamma_{\alpha\gamma}^{\gamma})\bigr),
\end{equation}
where the variation of the term $g^{\alpha\beta}(\delta\Gamma_{\beta\alpha}^{\sigma}) - g^{\alpha\sigma}
(\delta\Gamma_{\alpha\gamma}^{\gamma})$ is given in \ref{Appendix A}.
With this result the variation of the Ricci scalar becomes
\begin{align}
\delta R &= \delta g^{\alpha\beta}R_{\alpha\beta} + \nabla_{\sigma
}\bigl(g^{\alpha\beta}(\delta\Gamma_{\beta\alpha}^{\sigma}) - g^{\alpha\sigma}
(\delta\Gamma_{\alpha\gamma}^{\gamma})\bigr),\nonumber \\
&= \delta g^{\alpha\beta}R_{\alpha\beta} + g_{\mu\nu}\nabla_{\sigma}\nabla^{\sigma}(\delta g^{\mu\nu}) - \nabla_{\sigma}\nabla_{\gamma}(\delta g^{\sigma\gamma}),\nonumber \\
&= \delta g^{\alpha\beta}R_{\alpha\beta} + g_{\alpha\beta}\square(\delta g^{\alpha\beta}) - \nabla_{\alpha}\nabla_{\beta}(\delta g^{\alpha\beta}).
\end{align}
Here we define $\square \equiv\nabla_{\sigma}\nabla^{\sigma}$ and relabeled some indices. Putting the previous results together in the variation of the modified action (\ref{varaccion2}):
\begin{align}\label{varaccionf}
\delta S_{met} &= \int_{\mathcal{V}} d^4x \, \bigl(f(R)\delta\sqrt{-g} + \sqrt{-g}\, f'(R) \delta R\bigr), \nonumber \\
&= \int_{\mathcal{V}} d^4x \, \biggl(-f(R)\frac{1}{2}\sqrt{-g}\, g_{\alpha\beta}\delta g^{\alpha\beta} + f'(R)\sqrt{-g}\Bigl(\delta g^{\alpha\beta}R_{\alpha\beta} + g_{\alpha\beta}\square(\delta g^{\alpha\beta}) - \nabla_{\alpha}\nabla_{\beta}(\delta g^{\alpha\beta})\Bigr)\biggr),\nonumber \\\
&= \int_{\mathcal{V}} d^4x \, \sqrt{-g} \biggl(f'(R)\Bigl(\delta g^{\alpha\beta}R_{\alpha\beta} + g_{\alpha\beta}\square(\delta g^{\alpha\beta}) - \nabla_{\alpha}\nabla_{\beta}(\delta g^{\alpha\beta})\Bigr)-f(R)\frac{1}{2}\, g_{\alpha\beta}\delta g^{\alpha\beta}\biggr).
\end{align}
Now we will consider the next integrals:
\begin{equation}\label{integrals}
\int_{\mathcal{V}} d^4x \, \sqrt{-g} f'(R)g_{\alpha\beta}\square(\delta g^{\alpha\beta}), \qquad \int_{\mathcal{V}} d^4x \, \sqrt{-g} f'(R)\nabla_{\alpha}\nabla_{\beta}(\delta g^{\alpha\beta}).
\end{equation}
We shall see that these integrals can be expressed differently performing integration by parts. For this we define the next quantities:
\begin{equation}\label{M}
M_{\tau} = f'(R)g_{\alpha\beta}\nabla_{\tau}(\delta g^{\alpha\beta}) - \delta g^{\alpha\beta} g_{\alpha\beta}\nabla_{\tau}(f'(R)),
\end{equation}
and
\begin{equation}\label{N}
N^{\sigma} = f'(R)\nabla_{\gamma}(\delta g^{\sigma\gamma}) - \delta g^{\sigma\gamma}\nabla_{\gamma}(f'(R)).
\end{equation}
The combination $g^{\sigma\tau}M_{\tau} + N^{\sigma}$ is
\begin{equation}
g^{\sigma\tau}M_{\tau} + N^{\sigma} = f'(R)g_{\alpha\beta}\nabla^{\sigma}(\delta g^{\alpha\beta}) - \delta g^{\alpha\beta} g_{\alpha\beta}\nabla^{\sigma}(f'(R)) +
f'(R)\nabla_{\gamma}(\delta g^{\sigma\gamma}) - \delta g^{\sigma\gamma}\nabla_{\gamma}(f'(R)),
\end{equation}
in the particular case $f(R) = R$, the previous combination reduces to the expression (\ref{V}) with equation (\ref{vargamma}). The quantities $M_{\tau}$ and $N^{\sigma}$ allow us to write the variation of the bulk term (\ref{varaccionf}) in the following way (for details see \ref{Appendix B}):
\begin{multline}\label{varf1}
\delta S_{met} = \int_{\mathcal{V}} d^4x \, \sqrt{-g} \biggl(f'(R)R_{\alpha\beta} + g_{\alpha\beta}\square f'(R) - \nabla_{\alpha}\nabla_{\beta}f'(R)-f(R)\frac{1}{2}\, g_{\alpha\beta}\biggr)\delta g^{\alpha\beta}\\ + \oint_{\partial \mathcal{V}} d^{3}y\, \varepsilon\sqrt{|h|}n^{\tau}M_{\tau} + \oint_{\partial \mathcal{V}} d^{3}y\, \varepsilon\sqrt{|h|}n_{\sigma}N^{\sigma}.
\end{multline}
In the next section we will work out with the boundary contribution from (\ref{varf1}), and show how this terms cancel with the variations of the $S'_{GYH}$ action.
\subsection{Boundary terms in $f(R)$ gravity}
We express the quantities $M_{\tau}$ and $N^{\sigma}$ calculated in the boundary $\partial \mathcal{V}$. Is convenient to express them in function of the variations $\delta g_{\alpha\beta}$. Using the equation (\ref{varmet1}) in (\ref{M}) and (\ref{N}) yields :
\begin{equation}\label{M1}
M_{\tau} = -f'(R)g^{\alpha\beta}\nabla_{\tau}(\delta g_{\alpha\beta}) + g^{\alpha\beta}\delta g_{\alpha\beta} \nabla_{\tau}(f'(R)),
\end{equation}
and
\begin{equation}\label{N1}
N^{\sigma} = -f'(R)g^{\sigma\mu}g^{\gamma\nu}\nabla_{\gamma}(\delta g_{\mu\nu}) + g^{\sigma\mu}g^{\gamma\nu}\delta g_{\mu\nu}\nabla_{\gamma}(f'(R)).
\end{equation}
To evaluate this quantities in the boundary we use the fact that $\delta g_{\alpha\beta}|_{\partial \mathcal{V}}=\delta g^{\alpha\beta}|_{\partial \mathcal{V}}=0$, then the only terms not vanishing are the derivatives of $\delta g_{\alpha\beta}$ in the covariant derivatives. Hence we have
\begin{equation}\label{M2}
M_{\tau}\biggl|_{\partial \mathcal{V}} =-f'(R)g^{\alpha\beta}\partial_{\tau}(\delta g_{\alpha\beta}),
\end{equation}
and
\begin{equation}\label{M1}
N^{\sigma}\biggl|_{\partial \mathcal{V}} = -f'(R)g^{\sigma\mu}g^{\gamma\nu}\partial_{\gamma}(\delta g_{\mu\nu}),
\end{equation}
We now compute $n^{\tau}M_{\tau}\bigl|_{\partial \mathcal{V}}$ and $n_{\sigma}N^{\sigma}\bigl|_{\partial \mathcal{V}}$ which are the terms in the boundary integrals (\ref{varf1})
\begin{align}
n^{\tau}M_{\tau}\biggl|_{\partial \mathcal{V}} &= -f'(R)n^{\tau}(\varepsilon n^{\alpha}n^{\beta}+h^{\alpha\beta})\partial_{\tau}(\delta g_{\alpha\beta}), \nonumber\\
&= -f'(R)n^{\sigma}h^{\alpha\beta}\partial_{\sigma}(\delta g_{\alpha\beta}),
\end{align}
where we rename the dummy index $\tau$. In the other hand
\begin{align}
n_{\sigma}N^{\sigma}\biggl|_{\partial \mathcal{V}} &= -f'(R)n_{\sigma}(h^{\sigma\mu}+\varepsilon n^{\sigma}n^{\mu})(h^{\gamma\nu}+\varepsilon n^{\gamma}n^{\nu})\partial_{\gamma}(\delta g_{\mu\nu}), \nonumber\\
&= -f'(R)n^{\mu}(h^{\gamma\nu}+\varepsilon n^{\gamma}n^{\nu})\partial_{\gamma}(\delta g_{\mu\nu}), \nonumber\\
&= -f'(R)n^{\mu}h^{\gamma\nu}\partial_{\gamma}(\delta g_{\mu\nu}) \nonumber\\
&= 0,
\end{align}
where we have used that $n_{\sigma}h^{\sigma\mu}=0$, $\varepsilon^2=1$ and the fact that de tangential derivative $h^{\gamma\nu}\partial_{\gamma}(\delta g_{\mu\nu})$ vanishes. With this results the variation of the action $S_{met}$ becomes:
\begin{multline}\label{varf2}
\delta S_{met} = \int_{\mathcal{V}} d^4x \, \sqrt{-g} \biggl(f'(R)R_{\alpha\beta} + g_{\alpha\beta}\square f'(R) - \nabla_{\alpha}\nabla_{\beta}f'(R)-f(R)\frac{1}{2}\, g_{\alpha\beta}\biggr)\delta g^{\alpha\beta}\\ -\oint_{\partial \mathcal{V}} d^{3}y\, \varepsilon\sqrt{|h|}f'(R)n^{\sigma}h^{\alpha\beta}\partial_{\sigma}(\delta g_{\alpha\beta}).
\end{multline}
We proceed with the boundary term $S'_{GYH}$ in the total action. The variation of this term gives
\begin{align}
\delta S'_{GYH} &= 2\oint_{\partial \mathcal{V}} d^3y \, \varepsilon\sqrt{|h|}\bigl(\delta f'(R)K + f'(R)\delta K\bigr), \nonumber \\
&= 2\oint_{\partial \mathcal{V}} d^3y \, \varepsilon\sqrt{|h|}\bigl(f''(R)\delta R\, K + f'(R)\delta K\bigr).
\end{align}
Using the expression for the variation of $K$, equation (\ref{deltaK}), we can write\\
\begin{align}\label{varfront}
\delta S'_{GYH} &= 2\oint_{\partial \mathcal{V}} d^3y \, \varepsilon\sqrt{|h|}\biggl(f''(R)\delta R\, K + \frac{1}{2}f'(R)h^{\alpha\beta}\partial_{\sigma}(\delta g_{\beta\alpha})n^{\sigma}\biggr), \nonumber \\
&= 2\oint_{\partial \mathcal{V}} d^3y \, \varepsilon\sqrt{|h|}f''(R)\delta R\, K + \oint_{\partial \mathcal{V}} d^3y \, \varepsilon\sqrt{|h|}f'(R)h^{\alpha\beta}\partial_{\sigma}(\delta g_{\beta\alpha})n^{\sigma}.
\end{align}
We see that the second term in (\ref{varfront}) cancels the boundary term in the variation (\ref{varf2}), and in addition we need to impose $\delta R = 0$ in the boundary. Similar argument is given in \cite{Dyer}.\\\\
Finally, with the variation of the matter action, given in (\ref{variacionener}), the total variation of the action of modified $f(R)$ gravity is:\\
\begin{multline}
\delta S_{mod} = \frac{1}{2\kappa}\int_{\mathcal{V}} d^4x \, \sqrt{-g} \biggl(f'(R)R_{\alpha\beta} + g_{\alpha\beta}\square f'(R) - \nabla_{\alpha}\nabla_{\beta}f'(R)-\frac{1}{2}f(R)\, g_{\alpha\beta}\biggr)\delta g^{\alpha\beta}\\ -\frac{1}{2}\int_{\mathcal{V}} d^4x\,\sqrt{-g} T_{\alpha\beta}\delta g^{\alpha\beta}.
\end{multline}
Imposing that this variation becomes stationary we have:\\
\begin{equation}\label{variacionfinal}
\frac{1}{\sqrt{-g}}\frac{\delta S_{mod}}{\delta g^{\alpha\beta}} = 0  \Longrightarrow f'(R)R_{\alpha\beta} + g_{\alpha\beta}\square f'(R) - \nabla_{\alpha}\nabla_{\beta}f'(R)-\frac{1}{2}f(R)\, g_{\alpha\beta} = \kappa T_{\alpha\beta},
\end{equation}\\
which are the field equations in the metric formalism of $f(R)$ gravity.
\section{Conclusions}
We have obtained the field equations in the metric formalism of $f(R)$ gravity by using the direct results from variational principles.
The modified action in the metric formalism of $f(R)$ gravity plus a Gibbons-York-Hawking like boundary term must be written as:
\begin{equation}
S_{mod} = \frac{1}{2\kappa}\biggl[\int_{\mathcal{V}} d^4x\, \sqrt{-g}\Bigl(f(R) + 2\kappa\mathcal{L}_M[g_{\alpha\beta},\psi]\Bigl)+ 2\oint_{\partial \mathcal{V}} d^3y \, \varepsilon\sqrt{|h|} f'(R)K\biggr],
\end{equation}
with $f'(R) = df(R)/dR$ and $\mathcal{L}_M$ the lagrangian associated with all the matter fields. From the quantities $M_{\sigma}$ and $N^{\sigma}$, defined in (\ref{M}) and (\ref{N}) respectively, we recovered GR plus Gibbons-York-Hawking boundary term in the particular case $f(R)=R$. We see that including the boundary term, we have a well behaved mathematical problem setting both, $\delta g_{\alpha\beta}=0$ and $\delta R = 0$ in $\partial \mathcal{V}$.\\\\\\\\\\
\textbf{\large{Acknowledgements}:} The authors are grateful with the Observatorio Astron\'omico Nacional, Bogot\'a, Colombia, where this paper was carried out. A. Guarnizo acknowledges  the financial support by the Programa de Becas para Estudiantes Sobresalientes de Posgrado, Universidad Nacional de Colombia.

\appendix
\section{Evaluation of the term $g^{\alpha\beta}(\delta\Gamma_{\beta\alpha}^{\sigma}) - g^{\alpha\sigma}
(\delta\Gamma_{\alpha\gamma}^{\gamma})$}\label{Appendix A}
We already have calculated the variation $\delta \Gamma_{\beta\alpha}^{\sigma}$:
\begin{equation}
\delta \Gamma_{\beta\alpha}^{\sigma} = \frac{1}{2}\delta g^{\sigma\gamma}\bigl[\partial_{\beta}g_{\gamma\alpha} + \partial_{\alpha}g_{\gamma\beta} - \partial_{\gamma}g_{\beta\alpha}\bigr] + \frac{1}{2}g^{\sigma\gamma}\bigl[\partial_{\beta}(\delta g_{\gamma\alpha}) + \partial_{\alpha}(\delta g_{\gamma\beta}) -  \partial_{\gamma}(\delta g_{\beta\alpha})\bigr],
\end{equation}
writing the partial derivatives for the metric variations with the expression for the covariant derivative:
\begin{equation}
\nabla_{\gamma}\delta g_{\alpha\beta} = \partial_{\gamma}\delta g_{\alpha\beta} - \Gamma_{\gamma\alpha}^{\sigma}\delta g_{\sigma\beta} - \Gamma_{\gamma\beta}^{\sigma}\delta g_{\alpha\sigma},
\end{equation}
and also using that we are working in a torsion-free manifold i.e., the symmetry in the Christoffel symbol $\Gamma_{\beta\gamma}^{\alpha}=\Gamma_{\gamma\beta}^{\alpha}$, we can write:
\begin{align}
\delta \Gamma_{\beta\alpha}^{\sigma} &= \frac{1}{2}\delta g^{\sigma\gamma}\bigl[\partial_{\beta}g_{\gamma\alpha} + \partial_{\alpha}g_{\gamma\beta} - \partial_{\gamma}g_{\beta\alpha}\bigr] + \frac{1}{2}g^{\sigma\gamma}\bigl[\nabla_{\beta}(\delta g_{\gamma\alpha})+ \nabla_{\alpha}(\delta g_{\gamma\beta})-  \nabla_{\gamma}(\delta g_{\beta\alpha})  + \Gamma_{\beta\alpha}^{\lambda} \delta g_{\gamma\lambda} + \Gamma_{\alpha\beta}^{\lambda} \delta g_{\lambda\gamma} \bigr],\nonumber \\
&= \frac{1}{2}\delta g^{\sigma\gamma}\bigl[\partial_{\beta}g_{\gamma\alpha} + \partial_{\alpha}g_{\gamma\beta} - \partial_{\gamma}g_{\beta\alpha}\bigr] + g^{\sigma\gamma}\Gamma_{\beta\alpha}^{\lambda}\delta g_{\gamma\lambda}+ \frac{1}{2}g^{\sigma\gamma}\bigl[\nabla_{\beta}(\delta g_{\gamma\alpha})+ \nabla_{\alpha}(\delta g_{\gamma\beta})-  \nabla_{\gamma}(\delta g_{\beta\alpha}) \bigr],
\end{align}
using equation (\ref{varmet1}) in the second term:
\begin{align}
\delta \Gamma_{\beta\alpha}^{\sigma} &= \frac{1}{2}\delta g^{\sigma\gamma}\bigl[\partial_{\beta}g_{\gamma\alpha} + \partial_{\alpha}g_{\gamma\beta} - \partial_{\gamma}g_{\beta\alpha}\bigr] - \delta g^{\mu\nu}g^{\sigma\gamma}g_{\gamma\mu}g_{\lambda\nu}\Gamma_{\beta\alpha}^{\lambda}+ \frac{1}{2}g^{\sigma\gamma}\bigl[\nabla_{\beta}(\delta g_{\gamma\alpha})+ \nabla_{\alpha}(\delta g_{\gamma\beta})-  \nabla_{\gamma}(\delta g_{\beta\alpha}) \bigr],\nonumber \\
&= \delta g^{\sigma\nu}g_{\lambda\nu}\Gamma_{\beta\alpha}^{\lambda} - \delta g^{\mu\nu}\delta_{\mu}^{\sigma}g_{\lambda\nu}\Gamma_{\beta\alpha}^{\lambda} + \frac{1}{2}g^{\sigma\gamma}\bigl[\nabla_{\beta}(\delta g_{\gamma\alpha})+ \nabla_{\alpha}(\delta g_{\gamma\beta})-  \nabla_{\gamma}(\delta g_{\beta\alpha}) \bigr],\nonumber \\
&= \delta g^{\sigma\nu}g_{\lambda\nu}\Gamma_{\beta\alpha}^{\lambda} - \delta g^{\sigma\nu}g_{\lambda\nu}\Gamma_{\beta\alpha}^{\lambda} + \frac{1}{2}g^{\sigma\gamma}\bigl[\nabla_{\beta}(\delta g_{\gamma\alpha})+ \nabla_{\alpha}(\delta g_{\gamma\beta})-  \nabla_{\gamma}(\delta g_{\beta\alpha}) \bigr].
\end{align}
Then we have
\begin{equation}
\delta \Gamma_{\beta\alpha}^{\sigma} = \frac{1}{2}g^{\sigma\gamma}\bigl[\nabla_{\beta}(\delta g_{\alpha\gamma}) + \nabla_{\alpha}(\delta g_{\beta\gamma}) - \nabla_{\gamma}(\delta g_{\beta\alpha})\bigr],
\end{equation}
and similarly
\begin{equation}
\delta \Gamma_{\alpha\gamma}^{\gamma} = \frac{1}{2}g^{\sigma\gamma}\bigl[\nabla_{\alpha}(\delta g_{\sigma\gamma})\bigr].
\end{equation}
However it is convenient to express the previous result in function of the variations $\delta g^{\alpha\beta}$, we again use (\ref{varmet1}):
\begin{align}
\delta \Gamma_{\beta\alpha}^{\sigma} &= \frac{1}{2}g^{\sigma\gamma}\bigl[\nabla_{\beta}(- g_{\alpha\mu}g_{\gamma\nu}\delta g^{\mu\nu}) + \nabla_{\alpha}(- g_{\beta\mu}g_{\gamma\nu}\delta g^{\mu\nu}) - \nabla_{\gamma}(- g_{\beta\mu}g_{\alpha\nu}\delta g^{\mu\nu})\bigr],\nonumber \\
&= -\frac{1}{2}g^{\sigma\gamma}\bigl[g_{\alpha\mu}g_{\gamma\nu}\nabla_{\beta}(\delta g^{\mu\nu}) + g_{\beta\mu}g_{\gamma\nu}\nabla_{\alpha}(\delta g^{\mu\nu}) - g_{\beta\mu}g_{\alpha\nu}\nabla_{\gamma}(\delta g^{\mu\nu})\bigr],\nonumber \\
&= -\frac{1}{2}\bigl[\delta_{\nu}^{\sigma}g_{\alpha\mu}\nabla_{\beta}(\delta g^{\mu\nu}) + \delta_{\nu}^{\sigma}g_{\beta\mu}\nabla_{\alpha}(\delta g^{\mu\nu}) - g_{\beta\mu}g_{\alpha\nu}g^{\gamma\sigma}\nabla_{\gamma}(\delta g^{\mu\nu})\bigr],\nonumber \\
&= -\frac{1}{2}\bigl[g_{\alpha\gamma}\nabla_{\beta}(\delta g^{\sigma\gamma}) + g_{\beta\gamma}\nabla_{\alpha}(\delta g^{\sigma\gamma}) - g_{\beta\mu}g_{\alpha\nu}\nabla^{\sigma}(\delta g^{\mu\nu})\bigr],
\end{align}
where we write $\nabla^{\sigma} = g^{\sigma\gamma}\nabla_{\gamma}$. In a similar way:
\begin{equation}
\delta \Gamma_{\alpha\gamma}^{\gamma} = -\frac{1}{2}g_{\mu\nu}\nabla_{\alpha}(\delta g^{\mu\nu}).
\end{equation}
Now we compute the term $g^{\alpha\beta}(\delta\Gamma_{\beta\alpha}^{\sigma}) - g^{\alpha\sigma}
(\delta\Gamma_{\alpha\gamma}^{\gamma})$
\begin{align}
g^{\alpha\beta}(\delta\Gamma_{\beta\alpha}^{\sigma}) - g^{\alpha\sigma}
(\delta\Gamma_{\alpha\gamma}^{\gamma}) =& -\frac{1}{2}\Bigl(\bigl[g^{\alpha\beta}g_{\alpha\gamma}\nabla_{\beta}(\delta g^{\sigma\gamma}) + g^{\alpha\beta}g_{\beta\gamma}\nabla_{\alpha}(\delta g^{\sigma\gamma}) - g^{\alpha\beta}g_{\beta\mu}g_{\alpha\nu}\nabla^{\sigma}(\delta g^{\mu\nu})\bigr]\\
&-  \bigl[g^{\alpha\sigma}g_{\mu\nu}\nabla_{\alpha}(\delta g^{\mu\nu})\bigr]\Bigr),\nonumber \\
=& -\frac{1}{2}\Bigl(\bigl[\delta_{\gamma}^{\beta}\nabla_{\beta}(\delta g^{\sigma\gamma}) + \delta_{\gamma}^{\alpha}\nabla_{\alpha}(\delta g^{\sigma\gamma}) - \delta_{\mu}^{\alpha}g_{\alpha\nu}\nabla^{\sigma}(\delta g^{\mu\nu})\bigr]- \bigl[g_{\mu\nu}g^{\alpha\sigma}\nabla_{\alpha}(\delta g^{\mu\nu})\bigr]\Bigr),\nonumber \\
=& -\frac{1}{2}\Bigl(\bigl[\nabla_{\gamma}(\delta g^{\sigma\gamma}) + \nabla_{\gamma}(\delta g^{\sigma\gamma}) - g_{\mu\nu}\nabla^{\sigma}(\delta g^{\mu\nu})\bigr]- \bigl[g_{\mu\nu}\nabla^{\sigma}(\delta g^{\mu\nu})\bigr]\Bigr),\nonumber \\
=& -\frac{1}{2}\Bigl(2\nabla_{\gamma}(\delta g^{\sigma\gamma}) - 2g_{\mu\nu}\nabla^{\sigma}(\delta g^{\mu\nu})\Bigr),
\end{align}
then we have,
\begin{equation}\label{vargamma}
g^{\alpha\beta}(\delta\Gamma_{\beta\alpha}^{\sigma}) - g^{\alpha\sigma}
(\delta\Gamma_{\alpha\gamma}^{\gamma}) = g_{\mu\nu}\nabla^{\sigma}(\delta g^{\mu\nu})-\nabla_{\gamma}(\delta g^{\sigma\gamma}).
\end{equation}
\section{Integrals with $M_{\tau}$ and $N^{\sigma}$ }\label{Appendix B}
Taking the covariant derivative in $M_{\sigma}$:
\begin{align}
\nabla^{\tau}M_{\tau} &= \nabla^{\tau}\bigl(f'(R)g_{\alpha\beta}\nabla_{\tau}(\delta g^{\alpha\beta})\bigr) - \nabla^{\tau}\bigl(\delta g^{\alpha\beta} g_{\alpha\beta}\nabla_{\tau}(f'(R))\bigr),\nonumber \\
&= \nabla^{\tau}(f'(R))g_{\alpha\beta}\nabla_{\tau}(\delta g^{\alpha\beta}) + f'(R)g_{\alpha\beta}\square(\delta g^{\alpha\beta})  - \nabla^{\tau}(\delta g^{\alpha\beta}) g_{\alpha\beta}\nabla_{\tau}(f'(R))-\delta g^{\alpha\beta} g_{\alpha\beta}\square(f'(R)),\nonumber \\
&= f'(R)g_{\alpha\beta}\square(\delta g^{\alpha\beta})-\delta g^{\alpha\beta} g_{\alpha\beta}\square(f'(R)).
\end{align}
Here we have used the metric compatibility $\nabla^{\tau}g_{\alpha\beta}=0$, integrating this expression
\begin{equation}
\int_{\mathcal{V}} d^4x \, \sqrt{-g}\nabla^{\tau}M_{\tau} = \int_{\mathcal{V}} d^4x \, \sqrt{-g}f'(R)g_{\alpha\beta}\square(\delta g^{\alpha\beta})-\int_{\mathcal{V}} d^4x \, \sqrt{-g}\delta g^{\alpha\beta} g_{\alpha\beta}\square(f'(R)),
\end{equation}
using again the Gauss-Stokes theorem (\ref{gauss}), the first integral can be written as a boundary term:\\
\begin{equation}
\oint_{\partial \mathcal{V}} d^{3}y\, \varepsilon\sqrt{|h|}n^{\tau}M_{\tau} = \int_{\mathcal{V}} d^4x \, \sqrt{-g}f'(R)g_{\alpha\beta}\square(\delta g^{\alpha\beta})\bigr)-\int_{\mathcal{V}} d^4x \, \sqrt{-g}\delta g^{\alpha\beta} g_{\alpha\beta}\square(f'(R)),
\end{equation}
then we can write:
\begin{equation}
\int_{\mathcal{V}} d^4x \, \sqrt{-g}f'(R)g_{\alpha\beta}\square(\delta g^{\alpha\beta})=\int_{\mathcal{V}} d^4x \, \sqrt{-g}\delta g^{\alpha\beta} g_{\alpha\beta}\square(f'(R)) + \oint_{\partial \mathcal{V}} d^{3}y\, \varepsilon\sqrt{|h|}n^{\tau}M_{\tau}.
\end{equation}
In a similar way, taking the covariant derivative of $N^{\sigma}$:
\begin{align}
\nabla_{\sigma}N^{\sigma} &= \nabla_{\sigma}\bigl(f'(R)\nabla_{\gamma}(\delta g^{\sigma\gamma})\bigr) - \nabla_{\sigma}\bigl(\delta g^{\sigma\gamma}\nabla_{\gamma}(f'(R))\bigr),\nonumber \\
&= \nabla_{\sigma}(f'(R))\nabla_{\gamma}(\delta g^{\sigma\gamma})+ f'(R)\nabla_{\sigma}\nabla_{\gamma}(\delta g^{\sigma\gamma}) - \nabla_{\sigma}(\delta g^{\sigma\gamma})\nabla_{\gamma}(f'(R))-\delta g^{\sigma\gamma}\nabla_{\sigma}\nabla_{\gamma}(f'(R)),\nonumber \\
&= f'(R)\nabla_{\sigma}\nabla_{\beta}(\delta g^{\sigma\beta}) -\delta g^{\sigma\beta}\nabla_{\sigma}\nabla_{\beta}(f'(R)),
\end{align}
integrating:
\begin{equation}
\int_{\mathcal{V}} d^4x \, \sqrt{-g}\nabla_{\sigma}N^{\sigma} = \int_{\mathcal{V}} d^4x \, \sqrt{-g}f'(R)\nabla_{\sigma}\nabla_{\beta}(\delta g^{\sigma\beta}) -\int_{\mathcal{V}} d^4x \, \sqrt{-g}\delta g^{\sigma\beta}\nabla_{\sigma}\nabla_{\beta}(f'(R)),
\end{equation}
using again the Gauss-Stokes theorem we can write:
\begin{equation}
\int_{\mathcal{V}} d^4x \, \sqrt{-g}f'(R)\nabla_{\sigma}\nabla_{\beta}(\delta g^{\sigma\beta}) =\int_{\mathcal{V}} d^4x \, \sqrt{-g}\delta g^{\sigma\beta}\nabla_{\sigma}\nabla_{\beta}(f'(R))+\oint_{\partial \mathcal{V}} d^{3}y\, \varepsilon\sqrt{|h|}n_{\sigma}N^{\sigma}.
\end{equation}

\end{document}